\documentclass[12pt]{article}
\usepackage{amsmath,amssymb}
\usepackage{latexsym}

\newcommand{\cm}{\operatorname{cm}}
\newcommand{\lbar}{\mbox{$\lambda$\hspace*{-4.5pt}\raisebox{3pt}{-}}}
\newcommand{\gm}{\operatorname{gm}}

\begin{document}

\title{Nonlocal Electrodynamics of Linearly Accelerated Systems}

\author{Bahram Mashhoon\\Department of Physics and Astronomy\\University of
Missouri-Columbia\\Columbia, Missouri 65211, USA }

\maketitle

\begin{abstract} The measurement of an electromagnetic radiation field by a linearly
accelerated observer is discussed. The nonlocality of this process is emphasized. The nonlocal
theory of accelerated observers is briefly described and the
consequences of this theory are illustrated using a concrete example involving the measurement of an
 incident pulse of radiation by an observer that experiences uniform acceleration during a
limited interval of time.\end{abstract}

\setlength{\baselineskip}{24pt}

\section{Introduction}\label{s1}

Imagine an inertial frame in Minkowski spacetime in which the fundamental static inertial observers
measure the electric field ${\bf E}(t,{\bf x})$ and magnetic field ${\bf B}(t,{\bf x})$. The fields
measured by an inertial observer moving with uniform velocity ${\bf v}$ are then given by (see, e.g.
\cite{1})
\begin{align}\label{eq1} {\bf E}'_\parallel &={\bf E}_\parallel,& {\bf B}'_\parallel &={\bf
B}_\parallel,\\
\label{eq2}{\bf E}'_\bot &= \gamma \left( {\bf E}+\frac{{\bf v}}{c}\times {\bf B}\right)_\bot ,&
{\bf B}'_\bot & =\gamma \left({\bf B}-\frac{{\bf v}}{c}\times {\bf E}\right)_\bot,\end{align}
where $\gamma $ is the Lorentz factor of the observer. Equations \eqref{eq1} and \eqref{eq2} are
obtained from the fact that Maxwell's equations are invariant under Lorentz transformations. More
explicitly, let us define the Faraday tensor $F_{\mu\nu}\to ({\bf E,B})$ such that $F_{0i}=-E_i$ and
$F_{ij}=\epsilon_{ijk}B^k$. Greek indices run from $0$ to $3$ and Latin indices run from $1$ to $3$.
Under a Poincar\'e transformation $x^{'\mu}=L^\mu_{\;\; \nu}x^\nu+\alpha^\mu$, where $L^\mu_{\;\;\nu}$ is a
Lorentz matrix and $\alpha^\mu$ represents a spacetime translation,
\begin{equation}\label{eq3} F^{'\mu\nu}({\bf x}')=\frac{\partial x^{'\mu}}{\partial
x^\rho}\frac{\partial x^{'\nu}}{
\partial x^\sigma}F^{\rho\sigma}({\bf x}).\end{equation}
In component form, this expression reduces to equations \eqref{eq1} and \eqref{eq2}.

The basic laws of microphysics have been formulated with respect to ideal inertial observers. However,
all actual observers are accelerated. It is therefore necessary to determine what accelerated observers
measure. The term ``observer" is employed in this paper in an extended sense; for instance, an observer
could be an ideal measuring device. In the theory of measurement, it is important to distinguish the
practical aspects of a measurement from its basic theoretical aspects. Clearly a measuring device has
limitations due to the nature of its construction as well as its modes of operation. These limitations
need to be taken into account when using such a device in the laboratory. On the other hand, what is at
issue here is our theoretical expectation of an ideal device of this kind when it is accelerated. To
make contact with the basic laws of microphysics, the accelerated device must be related via a
theoretical postulate to ideal inertial devices. The postulate that is employed in the standard theory
of relativity is the hypothesis of locality \cite{2}-\cite{6}, namely, the assumption that the accelerated
observer is locally equivalent to an otherwise identical momentarily comoving inertial observer. Along
the worldline, the accelerated observer passes through a continuous infinity of such hypothetical
momentarily comoving inertial observers. To implement this hypothesis of locality, one must therefore
consider a class of Lorentz transformations between the global background inertial frame and the
inertial frames of the momentarily comoving inertial observers. It is convenient, however, to adopt a
different, but physically equivalent, approach based on the use of orthonormal tetrads.

Each inertial observer is endowed with an orthonormal tetrad $\lambda^\mu_{\;\;(\alpha )}$, where
$\lambda^\mu_{\;\;(0)}$ is the temporal axis of the observer and $\lambda^\mu_{\;\;(i)}$, $i=1,2,3$, are
the spatial axes of the observer. The measurement of the electromagnetic field by the inertial observer
amounts to the projection of the Faraday tensor on the observer's tetrad
\begin{equation}\label{eq4} F_{(\alpha
)(\beta)}=F_{\mu\nu}\lambda^\mu_{\;\;(\alpha)}\lambda^\nu_{\;\;(\beta )}.\end{equation}
For the fundamental set of static inertial observers in the background global inertial frame,
$\lambda^\mu_{\;\; (\alpha )}=\delta^\mu_{\;\;\alpha}$ and hence from equation \eqref{eq4}, ${\bf E}$
and ${\bf B}$ have the interpretation of fields measured by these observers. For an inertial observer
moving with velocity ${\bf v}$, equation \eqref{eq4} reduces to equations \eqref{eq1} and \eqref{eq2},
as expected. To see this explicitly in the case of motion along the $z$-axis, we note that the tetrad of
such an inertial observer is
\begin{align}\label{eq5} \lambda^\mu_{\;\;(0)}&=\gamma (1,0,0,\beta ), &
\lambda^\mu _{\;\;(1)}&= (0,1,0,0),\notag\\
\lambda^\mu_{\;\;(2)}&=(0,0,1,0), &
\lambda^\mu_{\;\;(3)}&=\gamma (\beta ,0, 0,1),\end{align}
where $\beta =v/c$ and $\gamma =(1-\beta^2)^{-\frac{1}{2}}$ is the Lorentz factor. Equation \eqref{eq4}
then implies that
\begin{align}\label{eq6} E'_1&=\gamma (E_1-\beta B_2),& B'_1&=\gamma (B_1+\beta E_2),\\
\label{eq7}E'_2&=\gamma (E_2+\beta B_1),& B'_2&=\gamma (B_2-\beta E_1),\\
\label{eq8}E'_3&=E_3, & B'_3 &=B_3,\end{align}
in agreement with equations \eqref{eq1} and \eqref{eq2}.

It follows from the hypothesis of locality that an accelerated observer is also endowed with an
orthonormal tetrad $\lambda^\mu_{\;\;(\alpha)}(\tau )$, where $\tau$ is the proper time along its
worldline $x^\mu=x^\mu(\tau )$,
\begin{equation}\label{eq9} \tau =\int^t_0\sqrt{1-\beta^2}\; dt.\end{equation}
Here it has been assumed that $\tau =0$ at $t=0$. The hypothesis of
locality implies that $\tau$ is the time recorded by an ideal clock
comoving with the accelerated observer. The question then arises whether the field measured by
the accelerated observer is point by point given by equation \eqref{eq4} as required by the hypothesis
of locality. To answer this question, we must first study the limitations of the hypothesis of locality.

Consider, for instance, the measurement of the frequency of a plane monochromatic electromagnetic wave
of frequency $\omega$ and wave vector ${\bf k}$ by an observer moving with a velocity ${\bf v}(t)$.
Assuming that at each instant of time $t$, the accelerated observer is equivalent to an inertial
observer with the same instantaneous velocity, the invariance of the phase of the wave under Lorentz
transformations implies that the frequency $\omega '$ measured by the observer is given by the Doppler
effect 
\begin{equation}\label{eq10} \omega '=\frac{\omega -{\bf k}\cdot
{\bf v}(t)}{\sqrt{1-\frac{v^2}{c^2}}}.\end{equation}
This frequency depends on the state of motion of the observer; therefore, equation (10) makes physical
sense if the observer is able to measure the frequency $\omega '$ during a period of time in which ${\bf
v}(t)$ does not change appreciably. Since the observer needs to record at least a few periods of the
wave in order to measure its frequency, equation \eqref{eq10} is physically reasonable if
\begin{equation}\label{eq11} n\frac{2\pi}{\omega} \left|\frac{d\; {\bf v}(t)}{dt}\right| \ll |{\bf
v}(t)|,\end{equation}
where $n\sim 1$. 
Here $n$ is the number of cycles of the wave used to measure its
frequency. In general, the frequency can be determined more accurately
if more cycles are employed; for large $n$, however, one finds from
equation \eqref{eq11} that the acceleration must then be very small.  
Combining equation \eqref{eq11} with $|{\bf v}|<c$, we find that $\lbar \ll c^2/A$, where $\lbar
=\lambda /(2\pi)$ is the reduced wavelength of the radiation and $A=|d\; {\bf v}/dt|$ is the magnitude of
the three-dimensional acceleration vector of the observer. Thus the Doppler formula in general makes sense if the reduced
wavelength of the radiation is much smaller than the {\it acceleration length} $\mathcal{L}=c^2/A$ of
the observer. On the other hand, the instantaneous Lorentz
transformations may be employed in accordance with the hypothesis of
locality to assign
electric and magnetic fields to the accelerated observer. These fields may then be Fourier analyzed in
the local frame of the observer to determine the frequency spectrum. It is expected that this spectrum
would then reduce to the Doppler formula in the JWKB limit.

To summarize, for accelerated systems the standard Doppler formula $\omega (\tau
)=-k_\mu u^\mu (\tau )$,  where $k^\mu$
is the wave four-vector and $u^\mu$ is the four-velocity of the observer, will produce a result that in general
varies in time. If the frequency changes from one instant of time to
the next, the period of the incident wave must in general be
negligibly short in order for the Doppler formula to make physical sense;
that is, $\lambda/\mathcal{L}\to 0$.

As a phenomenological alternative, a time-dependent frequency can also be defined as the inverse proper
time between successive peaks \cite{7}-\cite{10}. This has the advantage of defining the frequency over
an extended period of time, the same way it would be measured in practice. Both of these approaches
invoke the hypothesis of locality.

These phenomenological approaches can be extended to a gravitational
field along the lines indicated in \cite{11new}.  A discussion of the various
limitations on frequency measurements in the Schwarzschild geometry is
contained in \cite{12new}. 

The hypothesis of locality originates from Newtonian mechanics, where the state of a particle is given
by its position and velocity; the accelerated observer therefore has the same state as the comoving inertial
observer, hence they are equivalent. This is analogous to approximating a curve with its tangent line at
a point.  In fact, if all physical phenomena
could be reduced to pointlike coincidences of classical particles and rays of radiation, then the
hypothesis of locality would be exactly valid. On the other hand, it is not possible to measure an electric or magnetic field
instantaneously, as emphasized by Bohr and Rosenfeld \cite{11,12}. This paper will pursue a
nonlocal approach \cite{13}-\cite{17} that involves in essence an
integral averaging of all momentarily
equivalent inertial observers for the duration of the acceleration as
explained in section~\ref{s2}.

The expected deviation from the hypothesis of locality can be expressed in terms of acceleration
lengths. In general, the acceleration of the observer is given by an antisymmetric acceleration tensor
$\phi_{(\alpha )(\beta )}$ defined by
\begin{equation}\label{eq12} \frac{d\lambda^\mu_{\;\;(\alpha)}}{d\tau}=\phi_{(\alpha)}^{\;\;\;\;(\beta
)}\lambda^\mu _{\;\;(\beta )}.\end{equation}
The translational acceleration $a^\mu=du^\mu /d\tau $
is a spacelike
vector $(u^\mu a_\mu=0)$ and can be expressed as
$a^\mu=g^i\lambda^\mu_{\;\; (i)}$, where $a^\mu a_\mu={\bf g}\cdot
{\bf g}=g^2(\tau )$. Here $g(\tau )\geq 0$ is the magnitude of the
translational acceleration. In analogy with the Faraday tensor, the acceleration tensor may be expressed
as $\phi _{(\alpha )(\beta )}\to (-{\bf g}(\tau ),\boldsymbol{\Omega}(\tau ))$, where $g_i(\tau )=\phi
_{(0)(i)}$ and $\Omega _i(\tau )=\frac{1}{2}\epsilon_{ijk}\phi ^{(j)(k)}$. Here $\boldsymbol{\Omega}$
denotes the frequency of rotation of the spatial frame of the accelerated observer with respect to a
nonrotating (i.e. Fermi-Walker transported) frame. The invariant acceleration scales are constructed from
the scalars $\phi_{(\alpha )(\beta)}$; for instance, the lengths $\mathcal{L}=c^2/g$ and $c/\Omega$ and
the corresponding acceleration times $c/g$ and $1/\Omega$ refer to the
translational and rotational accelerations of the observer, respectively. The intrinsic acceleration
scales determine the scale of variation of the state of the observer; therefore, the hypothesis of
locality is a valid approximation if the intrinsic scale of the phenomenon under observation is
negligibly small compared to the corresponding acceleration scale of the observer. Thus the deviation from the hypothesis of locality is
expected to be proportional to $\lbar /\mathcal{L}$, where $\lbar$ is the intrinsic lengthscale of the
phenomenon under observation. For Earth-based optical experiments in the laboratory, for example,
$\lambda\sim 5000\; \overset{\circ}{\text{A}}$ for visible light, while $c^2/g_\oplus \simeq 1 \text{ lyr}$ and
$c/\Omega_\oplus \simeq 28\text{ au}$; therefore, $\lambda /\mathcal{L}_\oplus $ is
$\lesssim 10^{-19}$ and any deviations from locality appear to be too small to be detectable at present. On
the other hand, the development of ultrahigh-power lasers during the past fifteen years \cite{18,19} may
change the observational situation and could lead to the measurement of deviations from locality.

Nonlocal effects may become detectable with the help of laser pulses that can induce linear electron
accelerations of order $10^{24}\cm/\text{s}^2$ using the chirped pulse amplification technique
\cite{18,19}. Moreover, Sauerbrey \cite{20} has employed such high-intensity femtosecond
lasers to impart linear accelerations of order $10^{21}\cm/\text{s}^2$ to small grains. A grain with a
macroscopic mass of $\sim 10^{-12}\gm$ more closely approximates a classical accelerated observer in the
sense employed in relativity theory \cite{21}.

The nonlocal theory of accelerated observers is discussed in
section~\ref{s2}. The linearly accerlerated observer under
consideration in this paper is described in section~\ref{s3}. The
nonlocal electromagnetic measurements of the observer are studied in
sections~\ref{s4} and \ref{s5}. Section~\ref{s6} contains a brief
discussion of our results.

\section{Nonlocality}\label{s2}

According to the hypothesis of locality, the electromagnetic radiation field measured by an accelerated
observer is given by
\begin{equation}\label{eq13} F_{(\alpha )(\beta )}(\tau
)=F_{\mu\nu}(t(\tau ),{\bf x}(\tau ))\lambda^\mu_{\;\;(\alpha)}(\tau
  )\lambda^\nu_{\;\;(\beta )}(\tau ),\end{equation}
which is the projection of the Faraday tensor onto the tetrad of the accelerated observer. To go beyond
the hypothesis  of locality, one must find a more general relationship between the measurements of the
accelerated observer $\mathcal{F}_{(\alpha )(\beta)}(\tau )$ and the infinite class of momentarily
comoving inertial observers $F_{(\alpha )(\beta )}(\tau )$. The most general linear relationship between
$\mathcal{F}_{(\alpha )(\beta )}(\tau )$ and $F_{(\alpha )(\beta)}(\tau )$ consistent with causality is
\cite{13}
\begin{equation}\label{eq14} \mathcal{F}_{(\alpha )(\beta )}(\tau )=F_{(\alpha )(\beta)}(\tau
)+\int^\tau_{\tau_0}K_{(\alpha )(\beta)}^{\;\;\;\;\;\;\;\; (\gamma )(\delta)}(\tau ,\tau ')F_{(\gamma
)(\delta )}(\tau ')d\tau ',\end{equation}
where $\tau_0$ is the instant at which the acceleration is turned on and $K_{(\alpha
)(\beta)}^{\;\;\;\;\;\;\;\;(\gamma )(\delta)}$ is a kernel that is expected to be proportional to the
acceleration of the observer. For a radiation field with $\lbar /\mathcal{L}\to 0$, the nonlocal part of
the ansatz \eqref{eq14} is expected to vanish. The nonlocal ansatz \eqref{eq14} deals only with
spacetime scalars and is thus manifestly invariant under inhomogeneous Lorentz transformations of the
background spacetime.

The nonlocal part in equation \eqref{eq14} has the form of an average over the past worldline of the
accelerated observer; in the JWKB limit, the nonlocal part disappears and we recover the hypothesis of
locality \eqref{eq13}. Equation~\eqref{eq14} expresses a Volterra integral equation of the second kind.
According to Volterra's theorem, the relationship between $\mathcal{F}_{(\alpha )(\beta)}$ and
$F_{(\alpha )(\beta)}$ is unique in the space of continuous functions \cite{22}. Volterra's theorem has
been extended to the Hilbert space of square-integrable functions by Tricomi \cite{23}. It is useful to
rewrite equations \eqref{eq13} and \eqref{eq14} in matrix form by replacing the Faraday tensors by
six-vectors consisting of electric and magnetic fields. Thus \eqref{eq13} may be re-expressed as $\hat{F}=\Lambda
F$, where $F_{(\alpha)(\beta)}\to \hat{F}$, $F_{\mu\nu}\to F$ and $\Lambda$ is a $6\times 6$ matrix
constructed from the local tetrad frame. It follows that equation \eqref{eq14} can be written as
\begin{equation}\label{eq15} \hat{\mathcal{F}}(\tau )=\hat{F}(\tau )+\int^\tau _{\tau_0}\hat{K}(\tau
,\tau ')\hat{F}(\tau ')d\tau ',\end{equation}
where $\hat{K}$ is a $6\times 6$ matrix.

To find the kernel in equation \eqref{eq15}, we assume that no accelerated observer can ever be comoving
with an electromagnetic radiation field. This extends to all observers an important consequence of
Lorentz invariance: a basic radiation field can never stand completely still with respect to any
inertial observer. That is, if the incident radiation is non-constant (as it must be) for inertial
observers, it will be non-constant for any accelerated observer. Equivalently, if an accelerated
observer measures a constant electromagnetic radiation field, i.e. $\hat{\mathcal{F}}(\tau )=\hat{F}(\tau_0)$ in
equation \eqref{eq15}, then the inertial observers must also measure constant fields, i.e. $F$ must be
constant as well. The Volterra-Tricomi uniqueness theorem then ensures
that our physical requirement is satisfied for any radiation field: a
variable field will never be constant for any observer. 
Inserting these conditions in equation \eqref{eq15}, we find
\begin{equation}\label{eq16} \Lambda (\tau_0)=\Lambda (\tau )+\int^\tau_{\tau_0}\hat{K}(\tau ,\tau ')\Lambda
(\tau ')d\tau ',\end{equation}
which may be used to determine the kernel on the basis of our physical postulate. However, equation
\eqref{eq16} is not sufficient to determine the kernel uniquely. A detailed examination of
the possible kernels \cite{14,15} has revealed that the only acceptable solution is the kinetic kernel
given by $\hat{K}(\tau ,\tau ')=k(\tau ')$. In this case, equation \eqref{eq16} immediately implies that
\begin{equation}\label{eq17}k(\eta )=-\frac{d\Lambda (\eta)}{d\eta}\Lambda^{-1}(\eta ).\end{equation}
This kernel is directly proportional to the observer's acceleration and vanishes when it is turned off.
Moreover, it is constant for the case of uniform acceleration. A nonlocal theory of accelerated
observers has been developed on the basis of this unique kernel \cite{13}-\cite{16} and nonlocal
Maxwell's equations have been discussed in \cite{17}.

It is important to recognize that the hypothesis of locality is nevertheless an integral part of the
nonlocal theory described here: in the eikonal limit $\lbar /\mathcal{L}\to 0$, the nonlocal theory
reduces to the standard theory based on the hypothesis of locality. This is analogous to the
correspondence between wave mechanics and classical mechanics. The idea of such a correspondence will be
employed throughout this paper; for instance, we will assume that accelerated observers can perform
spatial and temporal measurements that are essentially consistent with the hypothesis of locality (see
section~\ref{s3}). A nonlocal treatment will be required only if the wave phenomena involved are such
that $\lbar /\mathcal{L}$ is not negligibly small and hence cannot be ignored.

\section{Linearly accelerated systems}\label{s3}

In this paper we are interested in the electromagnetic measurements of a linearly accelerated observer.
For the sake of concreteness, we assume that the observer moves
uniformly with velocity ${\bf v}=v_0\hat{\bf z}$ in the background global inertial
frame with coordinates $x^\alpha =(t,x,y,z)$ according to $x=x_0$, $y=y_0$ and
$z=z_0+v_0t$ for $-\infty <t<0$ and at $t=0$ is forced to accelerate with acceleration
$g(\tau )>0$ along the positive $z$-direction. Henceforth we use units
such that $c=1$, unless specified otherwise. The observer
carries a natural orthonormal tetrad frame $\lambda^\mu_{\;\;(\alpha)}$ given by
\begin{align} \lambda^\mu_{\;\;(0)} &=
    (C,0,0,S), &
\lambda^\mu_{\;\;(1)} & =(0,1,0,0),\notag \\
\lambda^\mu_{\;\;(2)}&=(0,0,1,0), &
\lambda^\mu_{\;\;(3)}&=(S,0,0,C),\label{eq18}\end{align}
where $C=\cosh \theta$, $S=\sinh \theta$ and
\begin{equation}\label{eq19} \theta = \theta_0+u(\tau )\int^\tau_0g(\tau ')d\tau '.\end{equation}
Here $\tanh \theta_0=v_0$ and $u(\tau )$ is the unit step function such that $u(\tau )=1$ for $\tau >0$ and $u(\tau )=0$ for
$\tau <0$. The orthonormal frame of the observer is nonrotating, i.e. it is Fermi-Walker transported
along the worldline of the observer. We recall that a Fermi-Walker transported four-vector $v^\mu$
carried along a worldline $x^\mu(\tau )$ obeys the transport equation
\begin{equation}\label{eq20} \frac{dv^\mu}{d\tau }=(u^\mu a^\nu-u^\nu a^\mu)v_\nu ,\end{equation}
where $u^\mu =\lambda^\mu_{\;\;(0)}$ is the four-velocity and $a^\mu=du^\mu/d\tau$ is the
four-acceleration vector of the worldline. Thus each leg of the tetrad \eqref{eq18} satisfies equation
\eqref{eq20}; moreover, we note that $a_\mu \lambda^\mu_{\;\;(3)}=u(\tau )g(\tau )$, where we have taken
the signature of the Minkowski metric to be $+2$. The magnitude of the
four-acceleration vector $g$ is related to the magnitude of the
three-acceleration vector $A$ by $g=A\gamma^3$, where $\gamma$ is the
Lorentz factor.

We assume for the sake of simplicity that the acceleration of the observer $g(\tau )$ is uniform and
equal to $g_0$ for $0<\tau <\tau _f$, but vanishes otherwise. That is
\begin{equation}\label{eq21} g(\tau )=g_0[u(\tau )-u(\tau -\tau _f)].\end{equation}
Therefore, the observer starts from its rest position $(x_0,y_0,z_0)$ at $t=0$ and accelerates uniformly
according to
\begin{equation}\begin{split}\label{eq22} t&=\frac{1}{g_0}[\sinh (\theta_0 +g_0\tau
    )-\sinh \theta_0],\quad x=x_0,\quad y=y_0,\\
z&=z_0+\frac{1}{g_0}[\cosh (\theta_0 +g_0\tau )-\cosh \theta_0]\end{split}\end{equation}
until the time $t_f$, where $g_0t_f=\sinh (\theta_0+g_0\tau_f)-\sinh \theta_0$. For $t>t_f$, the observer moves uniformly with
speed $\beta_f=\tanh (\theta_0+g_0\tau_f)$,
\begin{equation}\label{eq23} x=x_0,\quad y=y_0,\quad z=z_f+\beta_f(t-t_f),\end{equation}
where $z_f$ is given by $g_0(z_f-z_0)=\cosh
(\theta_0+g_0\tau_f)-\cosh \theta_0$ using equation \eqref{eq22}. The proper time of
the observer for $t>t_f$ is given by
\begin{equation}\label{eq24} \tau =\tau_f+\frac{t-t_f}{\cosh (\theta_0+g_0\tau_f)}.\end{equation}

It follows from the results of the previous section that in this case
\begin{equation}\label{eq25} \Lambda =\begin{bmatrix} U & V\\ -V &
  U\end{bmatrix}, \quad U=\begin{bmatrix} C & 0 & 0\\ 0 & C & 0\\ 0 &
  0 &1\end{bmatrix},\quad V=SI_3,\end{equation}
where $I_i$, $(I_i)_{jk}=-\epsilon_{ijk}$, is a $3\times 3$ matrix
  proportional to the operator of infinitesimal rotations about the
  $x^i$-axis. According to the postulates of the nonlocal theory of
  accelerated observers, the fields as measured by the linearly
  accelerated observer are given by equation \eqref{eq14}, where the
  kernel, given by equation \eqref{eq17}, reduces in this case to
\begin{equation}\label{eq26} k(\tau )=-g(\tau )\begin{bmatrix} 0 & I_3\\
  -I_3 & 0\end{bmatrix}.\end{equation}

In the standard theory of relativity, the field as measured by the
accelerated observer is determined via the hypothesis of locality and
for the specific case of the linearly accelerated observer described
in this section
\begin{align}\label{eq27} E_{(1)}&=CE_1-SB_2, & \quad B_{(1)}&=CB_1+SE_2,\\
E_{(2)}&=CE_2+SB_1, & \quad B_{(2)}&=CB_2-SE_1,\label{eq28}\\
E_{(3)}&=E_3, &\quad B_{(3)}&=B_3, \label{eq29}\end{align}
where $(C,S)=(\cosh \theta_0,\sinh\theta_0)$ for $t<0$, $(C,S)=(\cosh
(\theta_0+g_0\tau ),\sinh (\theta_0+g_0\tau ))$ for $0\leq t\leq t_f$
and $(C,S)=(\cosh (\theta_0+g_0\tau_f), \sinh (\theta_0+g_0\tau _f))$
for $t>t_f$. However, according to the nonlocal theory the electric
field is given by
\begin{align}\label{eq30} \mathcal{E}_{(1)}&= E_{(1)}+u(\tau
  )\int^\tau_0g(\tau ')B_{(2)}(\tau ')d\tau ',\\
\label{eq31}\mathcal{E}_{(2)}&=E_{(2)}-u(\tau )\int^\tau_0g(\tau ')B_{(1)}(\tau
')d\tau ',\\
\label{eq32} \mathcal{E}_{(3)}&=E_{(3)}.\end{align}
Similarly, the nonlocal magnetic field is given by
\begin{align}\label{eq33} \mathcal{B}_{(1)}&=B_{(1)}-u(\tau
  )\int^\tau_0g(\tau ') E_{(2)}(\tau ')d\tau ',\\
\label{eq34} \mathcal{B}_{(2)}&= B_{(2)}+u(\tau )\int^\tau_0g(\tau
')E_{(1)}(\tau')d\tau ',\\
\label{eq35}\mathcal{B}_{(3)}&= B_{(3)}.\end{align}
It follows that the components of the electric and magnetic fields
parallel to the direction of motion of the observer remain the
same. Moreover, for $\tau >\tau_f$ the observer moves uniformly yet
its measurement of the electromagnetic field yields in addition to the
standard result a new constant component that is in effect the memory
of the past acceleration of the observer.

The measurements of the accelerated observer for the case of a
perpendicularly incident plane-polarized Gaussian pulse of
electromagnetic radiation were numerically investigated in \cite{24} on
the basis of the hypothesis of locality, namely, equations
\eqref{eq27}-\eqref{eq29}. The same situation has recently been
studied in connection with the {\em nonlocal} measurements of the
observer involving equations \eqref{eq30}-\eqref{eq35}; in fact, an
extension of the previous work \cite{24} and a detailed numerical
analysis of nonlocality in this case are contained in \cite{25}.

To illustrate further the physical consequences of the nonlocal
theory, two rather distinct situations will be explicitly worked out
in the rest of this paper. In the first case (``parallel incidence''),
discussed in the next section, the electromagnetic radiation
propagates along the $z$-axis. This situation is analogous to the
acceleration of grains by a high-intensity femtosecond laser pulse;
indeed, the results of this work should be compared and contrasted
with the analysis of the experimental situation presented in
\cite{20,21}. The second case (``perpendicular incidence''), discussed
in section~\ref{s5}, involves a pulse of plane-polarized radiation
that propagates along the $x$-axis; the nonlocal measurements of the
accelerated observer are compared with the standard theory. To
simplify the analysis, we take advantage of the fact that all of the
field operations considered in this work are linear. Therefore, it
suffices to focus attention on a generic Fourier component of the
incident pulse. Moreover, we use complex fields whose real parts
correspond to the measured fields. Thus to recover the actual
predictions of the theory, one must take the real part of the Fourier
sum of the results given in sections~\ref{s4} and \ref{s5}.

\section{Parallel incidence}\label{s4}

Imagine a plane monochromatic wave of frequency $\omega$ given in the
circular polarization basis by
\begin{equation}\label{eq36} {\bf E}=E_0\;{\bf e}_\pm \;e^{i\omega (z-t)},\quad
  {\bf B}=E_0\;{\bf b}_\pm \; e^{i\omega (z-t)},\end{equation}
where ${\bf e}_\pm =(\hat{\bf x}\pm i\hat{\bf y})/\sqrt{2}$, ${\bf
  b}_\pm =\mp i{\bf e}_\pm$ and $E_0(\omega)$ is the amplitude of the
wave. The wave propagates along the $z$-direction and at $t=0$
impinges on a grain (``observer'') at $z=z_0$ that accelerates along
the $z$-axis with initial velocity $v_0$. For $0\leq \tau \leq
\tau_f$, it follows from equation \eqref{eq22} that in the six-vector
notation of section~\ref{s2}
\begin{equation}\label{eq37} F(\tau )=E_0(\omega)\phi (\theta
  )\begin{bmatrix} {\bf e}_\pm \\{\bf
  b}_\pm\end{bmatrix},\end{equation}
where $\phi$ is given by
\begin{equation}\label{eq38} \phi (\theta)=e^{i\omega z_0}\exp \left[
  i\frac{\omega}{g_0}(e^{-\theta}-e^{-\theta_0})\right]\end{equation} and
  $\theta =\theta_0+g_0\tau$. Moreover, $\hat{F}=\Lambda F$ implies
  that in this case
\begin{equation}\label{eq39} \hat{F}=e^{-\theta}F.\end{equation}
This result together with the kernel \eqref{eq26} can be substituted
in equation~\eqref{eq15} with $\tau_0=0$ and after a simple
integration the nonlocal result is
\begin{equation}\label{eq40} \hat{\mathcal{F}}(\tau )=E_0\Phi (\tau
  )\begin{bmatrix} {\bf e}_\pm \\ {\bf b}_\pm
  \end{bmatrix},\end{equation}
where
\begin{equation}\label{eq41} \Phi (\tau )=e^{-\theta (\tau )}\phi
  (\theta)+i\frac{g_0}{\omega} [\phi (\theta )-\phi
  (\theta_0)].\end{equation}
It is important to note that the nonlocal theory, just as in the local
  case, does not introduce any coupling between the photon helicity
  and the acceleration of the observer. The nonlocal contribution to
  $\Phi$ is given by $i(g_0/\omega )[\phi (\theta )-\phi (\theta_0)]$
  in equation~\eqref{eq41}; as expected, it is proportional to $\lbar
  /\mathcal{L}_0=g_0/\omega$, which for the experiments described in
  \cite{20,21} is negligibly small and of the order of $10^{-5}$.

In connection with the treatment of \cite{20,21}, it is interesting to
note that in equation~\eqref{eq22}, one can expand in powers of
$g_0\tau \ll c$ to get
\begin{equation}\label{eq42} t=\gamma_0\tau \left(
  1+\frac{1}{2}\beta_0\frac{g_0\tau}{c}+\dotsb \right), \;\;
  z=z_0+\gamma_0\tau \left(v_0+\frac{1}{2}g_0\tau +\dotsb \right),\end{equation}
where $\beta_0=v_0/c$, $\gamma_0$ is the corresponding Lorentz factor
  and in equation~\eqref{eq38}
\begin{equation}\label{eq43}
  e^{-\theta_0}=\sqrt{\frac{1-\beta_0}{1+\beta_0}}.\end{equation}
It follows that
\begin{equation}\label{eq44} z=z_0+v_0t+\frac{1}{2}A_0t^2+\dotsb
  ,\end{equation}
where $A_0=g_0/\gamma_0^3$ is the magnitude of the initial
three-dimensional acceleration vector. Using these results, $\phi$ and
$\Phi$ in
equation~\eqref{eq41} can be expressed as
\begin{align}\label{eq45} \phi & =e^{i\frac{\omega}{c}\left(
    -ct+z_0+v_0t+\frac{1}{2}A_0t^2+\dotsb \right)},\notag \\
\Phi &= \left[ \sqrt{\frac{1-\beta_0}{1+\beta_0}} +
i\frac{g_0}{c\omega} + \frac{g_0}{c^2}\left( -ct
+v_0t+\frac{1}{2}A_0t^2 + \dots\right) \right]\phi -
i\frac{g_0}{c\omega} e^{i\frac{\omega}{c}z_0}\;, 
\end{align}
in general agreement with \cite{21}. One can thus draw the conclusion
that the nonlocal contribution to the field that sets the charged
particles in the grain in motion has an amplitude of order $10^{-5}$
compared to the standard local theory for the experiments reported in
\cite{20,21}. The reflected wave is therefore expected
to be affected at essentially the same negligibly small level of $\sim
10^{-5}$. It is this reflected wave that is detected and analyzed in
practice \cite{20,21}. The nonlocal theory in this case
would thus appear to be trivially consistent with the
results of experiments reported in \cite{20,21}, since the nonlocality
enters the analysis at the insignificant level of $g_0/\omega \sim 10^{-5}$.

Nonlocal effects may not be negligible in future experiments using
macrophysical accelerated systems \cite{20,21}. A remark is therefore in
order here regarding the relevance of our present  calculations  that
are based on a single plane wave. It should be emphasized that further
extensive calculations using wave packets would be necessary in order
that the nonlocal theory could be properly compared with experimental
data regarding accelerated plasmas generated by an incident
femtosecond laser pulse \cite{20,21}. What is measured in such experiments
is the shape of the reflected pulse \cite{20,21}; therefore, to determine
the anticipated contribution of nonlocality, one should consider an
incident pulse in the accelerated frame that is a Fourier sum of terms
of the form \eqref{eq41}. One must then take into account the interaction of
this incident radiation with the accelerated medium via an effective
reflectivity function as in \cite{20}. The theoretical determination of the
reflected pulse in the laboratory frame would then require an inverse
nonlocal transformation -- given in its general form by equation (19)
of \cite{17} -- and hence further detailed considerations that are beyond
the scope of the present paper.  

For $\tau \geq \tau_f$, the measurements of the observer are given by
equation~\eqref{eq40}, where $\Phi (\tau )$ for $\tau \geq \tau_f$ can be expressed as
\begin{equation}\label{eq46} \Phi (\tau )=e^{-\theta
    _f}\tilde{\phi}(\tau )+i \frac{g_0}{\omega}[\phi (\theta_f)-\phi
    (\theta_0)].\end{equation}
Here $\theta _f=\theta_0+g_0\tau_f$ and $\tilde{\phi}(\tau )$ is
    given by
\begin{equation}\label{eq47} \tilde{\phi}(\tau )=\exp \left\{
    i\omega \left[z_f-t_f-\sqrt{\frac{1-\beta_f}{1+\beta_f}}(\tau
    -\tau_f)\right]\right\}.\end{equation}
The second term in $\Phi (\tau )$ is the constant memory of the
    observer's past acceleration.

\section{Perpendicular incidence}\label{s5}

Consider a linearly-polarized plane monochromatic wave of frequency
$\omega$ propagating in the $x$-direction. The electric and magnetic
fields are given by
\begin{equation}\label{eq48} {\bf E}=E_0\; e^{i\omega (x-t)}\; \hat{\bf z},
  \quad {\bf B}=-E_0\; e^{i\omega (x-t)}\; \hat{\bf y}.\end{equation}
At $t=0$ the wave impinges upon a grain (``observer'') at
$(x_0,y_0,z_0)$ that has been prearranged to accelerate along the $z$-direction with initial
velocity $v_0$. To determine the field measured by the observer, it is
useful to assume that $0\leq \tau \leq \tau_f$ and define $\chi
(\theta)$,
\begin{equation}\label{eq49} \chi (\theta)=E_0\; e^{i\omega x_0}\exp
  \left[ -i\frac{\omega}{g_0}(\sinh \theta -\sinh \theta_0)\right]
  ,\end{equation}
where $\theta =\theta_0+g_0\tau$. The measured components of the
  electromagnetic field according to the nonlocal theory are then
\begin{equation}\label{eq50} \mathcal{E}_{(1)}=\sinh \theta \;\chi
  (\theta )-\int^\theta_{\theta_0}\cosh \theta '\; \chi (\theta ')\; d\theta
  ',\end{equation}
$\mathcal{E}_{(2)}=0$, $\mathcal{E}_{(3)}=\chi (\theta)$,
  $\mathcal{B}_{(1)}=\mathcal{B}_{(3)}=0$ and 
\begin{equation}\label{eq51} \mathcal{B}_{(2)}=-\cosh \theta \;\chi
  (\theta )+\int^\theta_{\theta_0}\sinh \theta '\;\chi (\theta ')\; d\theta
  '.\end{equation}
It is simple to work out the integral in equation~\eqref{eq50} and the
  result is
\begin{equation}\label{eq52} \mathcal{E}_{(1)}=\sinh \theta \; \chi
  (\theta )-i\frac{g_0}{\omega}[\chi (\theta )-\chi
  (\theta_0)],\end{equation}
where $\chi (\theta_0)=E_0\exp (i\omega x_0)$. The integral in
  equation~\eqref{eq51} can be evaluated by expanding the exponential
  function in $\chi (\theta ')$ in powers of $\sinh \theta '$ and then
  using formulas 2.412 on p. 93 of \cite{26}.

For $\tau \geq \tau_f$, the fields measured by the uniformly moving
observer are
\begin{equation}\label{eq53} \mathcal{E}_{(1)}=\sinh \theta_f
  \; \tilde{\chi}(\tau )-i\frac{g_0}{\omega }[\chi (\theta_f)-\chi
  (\theta_0)],\end{equation}
$\mathcal{E}_{(2)}=0$, $\mathcal{E}_{(3)}=\tilde{\chi}$,
  $\mathcal{B}_{(1)}=\mathcal{B}_{(3)}=0$ and
\begin{equation}\label{eq54} \mathcal{B}_{(2)}=-\cosh
  \theta_f\; \tilde{\chi}(\tau )+\int^{\theta_f}_{\theta_0}\sinh \theta
  '\; \chi (\theta ')\; d\theta '.\end{equation}
Here $\tilde{\chi}(\tau )$ is given by
\begin{equation}\label{eq55}\tilde{\chi}(\tau )=E_0\; e^{i\omega [x_0-t_f-\gamma
  _f(\tau -\tau_f)]},\end{equation}
where $\gamma_f=\cosh \theta_f$. The constant terms in
equations~\eqref{eq53} and \eqref{eq54} are the remnants of the
observer's accelerated history.

\section{Discussion}\label{s6}

We have examined some of the observational consequences of the nonlocal theory of accelerated observers for the case of
linearly accelerated systems. As in the standard local theory, the electromagnetic radiation field parallel to the direction
of motion remains unchanged and, moreover, there is no coupling between
the helicity of the radiation and the translational acceleration of
the observer \cite{29,30}. The observer's nonlocal determination of the
electromagnetic radiation field has been compared and contrasted with
the standard local ansatz. The results are consistent with a relative nonlocal contribution to the field given
essentially by the amplitude $\lbar /\mathcal{L}$, which turns out to
be negligibly small for the experiments reported in \cite{20,21}. Moreover, for an
observer that has resumed uniform motion, the constant electromagnetic
memory of past acceleration has been
 studied. The observation of such nonlocal effects may become possible with methods that use high-power laser
systems to generate large accelerations \cite{18}-\cite{21}.

\section*{Acknowledgments}

Thanks are due to C. Chicone and J. Hauck for helpful discussions.
I am grateful to G. Sch\"afer and R. Sauerbrey for their constructive comments.

\end{document}